\documentclass[11pt,twoside]{article}

\usepackage{asp2006}
\usepackage{epsf}
\usepackage{psfig}
\usepackage{lscape}

\begin{document}
\title{Massive Star Forming Regions: Turbulent Support or Global Collapse?}   %%% Fill in title
\author{E. V\'azquez-Semadeni,\altaffilmark{1} J. Ballesteros-Paredes,\altaffilmark{1} R. S. Klessen,\altaffilmark{2} A. K. Jappsen\altaffilmark{3}} %%% Fill in author names

\altaffiltext{1}{Centro de Radioastronom\'\i a y Astrof\'\i sica,
Universidad Nacional Aut\'onoma de M\'exico, Apdo. Postal 3-72, Morelia,
Michoac\'an, 58089, M\'exico}
\altaffiltext{2}{Zentrum f\"ur Astronomie der Universit\"at Heidelberg,
Institut f\"ur Theoretische Astrophysik, 69120 Heidelberg, Germany} 
\altaffiltext{3}{School of Physics \& Astronomy, Cardiff University,
Queens Buildings, The Parade, Cardiff CF24 3AA, UK}
%\altaffiltext{4}{School of Physics, University of Exeter, Exeter EX4
%4QL, UK}

\begin{abstract} %%% Abstract to run on from here.
We present preliminary numerical evidence that the physical conditions
in high-mass star forming regions can arise from global gravitational
infall, with the velocity dispersions being caused primarily by infall
motions rather than random turbulence. To this end, we study the clumps
and cores appearing in the region of central collapse in a numerical
simulation of the formation, evolution, and subsequent collapse of a
dense cloud out of a transonic compression in the diffuse atomic
ISM. The clumps have sizes $\sim 1$ pc, masses of several hundred
$M_\odot$, and three-dimensional velocity dispersions $\sim 3$ km
s$^{-1}$, in agreement with typical observed values for such
structures. The clumps break down into massive cores of sizes $\sim 0.1$
pc, densities $\sim 10^5$, masses 2-300 $M_\odot$, with distributions of
these quantities that peak at the same values as the massive core sample
in a recent survey of the Cygnus X molecular cloud complex.  Although
preliminary, these results suggest that high-mass star forming clumps
may be in a state of global gravitational collapse rather than in
equilibrium supported by strong turbulence.
\end{abstract}

%%% MAIN BODY OF TEXT GOES HERE. CONSULT "INSTRUCTIONS FOR AUTHORS USING
%%% LATEX2E MARKUP", SECTIONS 2.3-2.6 FOR HELP WITH EQUATIONS, FIGURES,
%%% AND TABLES.

\section{Introduction} \label{sec:intro}
High-mass star forming regions are characterized by more extreme
physical conditions than their low-mass counterparts
\citep[e.g.][]{GL99, Kurtz_etal00, Beuther_etal07}, having ``clumps'' of
sizes 0.2--0.5 pc, mean densities $n \sim 10^5$ cm$^{-3}$, masses
between 100 and 1000 $M_\odot$, and velocity dispersions ranging between
1.5 and 4 km s$^{-1}$. In turn, these clumps break down into even denser
``cores'' that are believed to be the immediate precursors of single or
gravitationally bound multiple massive protostars. The high velocity
dispersions of these clumps are generally interpreted as strong
turbulence that manages to support the clumps against gravity
\citep[e.g.,][]{GL99,MT03}. However, the notion of ``turbulent support''
is difficult to maintain at the scales of these cores. Turbulence is a
flow regime in which the largest velocity differences are associated
with the largest separations \citep[e.g.][]{Frisch95}, and moreover in
the case of supersonic turbulence the clumps are expected to be formed
by large-scale compressive motions, so that the turbulence is likely to
have a strong compressive component \citep{HF82, BVS99, BP_etal08,
VS_etal08}. So, it is difficult to imagine a turbulent velocity field
inside the clumps that is completely random in such a way as to only
provide support against gravity -- the compressive component may 
rather foster core contraction.
%not connected to the large-scale velocity
%field outside it. 
In addition, numerical simulations of cloud formation
in the diffuse atomic ISM \citep[e.g.][]{VPP95, VPP96, PVP95, BVS99,
BHV99, AH05, Heitsch_etal05,
Heitsch_etal06, VS_etal06, VS_etal07, HA07} and of star formation in
turbulent, self-gravitating clouds \citep[e.g.][]{KHM00, HMK01, BBB03,
VKSB05, VS_etal07} show that the velocity fields are organized at all
scales, exhibiting a continuity from the large scales outside the clumps
all the way to their interiors. 

The large-scale compressions can be of turbulent origin (e.g., passing
spiral arm shocks, supernova shells, or simply the general transonic
turbulence in the diffuse medium) or of gravitational origin
\citep[e.g., large-scale gravitational or magneto-gravitational
instabilities; ][]{Elm91, KOS02, FBK08}. The self-gravitating
simulations mentioned above exhibit gravitationally driven motions up to
the largest scales. In particular, \citet[hereafter
Paper I]{VS_etal07} have presented simulations of molecular cloud
formation by generic compressions in the diffuse atomic ISM and of its
subsequent collapse and star-forming stage, using the 
SPH/N-body code GADGET \citep[][]{Springel_etal01}, complemented with
parameterized heating and cooling taken from \citet{KI02} and a
sink-particle prescription \citep{Jappsen_etal05}. In these simulations,
the clouds that formed acquired their initial turbulence from
instabilities of the compressed layer \citep{Vishniac94, WF98, KI02,
Heitsch_etal05, Heitsch_etal06, VS_etal06}, but soon they became
gravitationally unstable and began contracting. This contraction phase
was nevertheless characterized by a virial-like energy balance with
$|E_{\rm grav}| \sim 2 E_{\rm kin}$, which was however due to the
gravitational contraction, not to virial equilibrium. During the
global contraction, clumps produced by the initial turbulence proceeded to
collapse on their own, forming what resembled low-mass star forming
regions. 

In Paper I, we speculated that the global contraction might be halted by
stellar energy feedback before the global collapse was
completed. However, here we forgo that speculation, and take the
simulation at face value, presenting a preliminary study of the physical
conditions in the region where the global collapse finally converges,
showing that they resemble the physical conditions of high-mass star
forming regions, thus suggesting that such regions may be in generalized
gravitational collapse rather than in a state of turbulent ``support''.

\section{The numerical model} \label{sec:model}

The simulation we consider is the one labeled L256$\Delta v$0.17 in
Paper I. We refer the reader to that paper for details. Here we just
mention that it is an SPH simulation with self-gravity, parameterized
heating and cooling implying a thermally bistable medium, using $3.24
\times 10^6$ particles, using sink particles, and initially set up to
produce a collision of streams of diffuse gas (at the same density as
their surroundings) that induces a transition to the cold, dense phase
and the generation of turbulence in the dense gas. The size of the
numerical box was 256 pc, and the inflow velocity of the colliding
streams was 1.25 times the sound speed in the ambient gas, which had $T
= 5000$ K and a mean density of $n = 1$ cm$^{-3}$. The turbulent cloud
eventually reached densities typical of molecular gas, and began to
contract gravitationally. The cloud had a flattened shape during most of
its evolution, and star formation began at $t \sim 17$ Myr in the
periphery of the cloud, where secondary compression produced by the gas
squirting off the collision site produced the highest initial densities.
Animations showing the large-scale evolution of this simulation can be
found in the electronic edition of Paper I.

\begin{figure}
\plotone{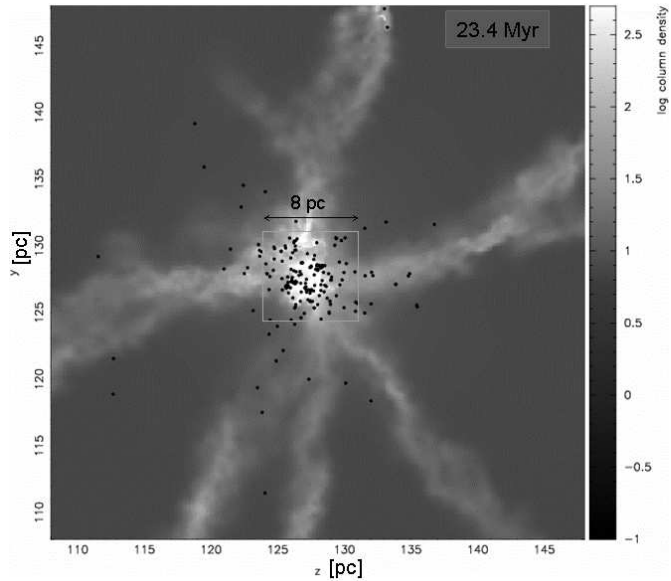}
\caption{Column density plot of the central 50 pc of this simulation in the
$y$-$z$ plane at $t=23.4$ Myr, integrating over the central 8 pc along
the $x$ direction. The box shows the region analyzed in \S
\ref{sec:phys_cond_ctr}. The dots show the stellar objects (sink
particles) already formed in the simulation by this time. Radial
filamentary streams of intermediate-density gas are seen to still be
accreting onto the central cloud.}
\label{fig1}
\end{figure}

By $t = 23.4$ Myr, the global collapse is completed, although the
residual turbulence causes the motions to have a random component, so
that the collapse center spans several parsecs across. Figure \ref{fig1}
shows a column density of the central 50 pc of this simulation in the
$y$-$z$ plane at $t=23.4$ Myr, integrating over the central 8 pc along
the $x$ direction. The dense cloud is seen near the center of the image,
with streams of gas still infalling onto it. A region 8 pc on a side
containing the cloud is indicated by the square, for which
animations can be found at {\tt
http://www.astrosmo.unam.mx/$\sim$e.vazquez/turbulence/movies.html}.
These animations show the evolution of the central 8 pc of
the simulation for $22.1 \le t \le 24.7$ Myr, showing a violent
collapsing evolution in which infalling clumps of gas interact but do
not entirely merge, but rather undergo shredding and distortion. It is
this violent region that we analyze in the next section.

\section{Physical conditions of clumps and cores in the collapse center}
\label{sec:phys_cond_ctr}

\begin{figure}
\plottwo{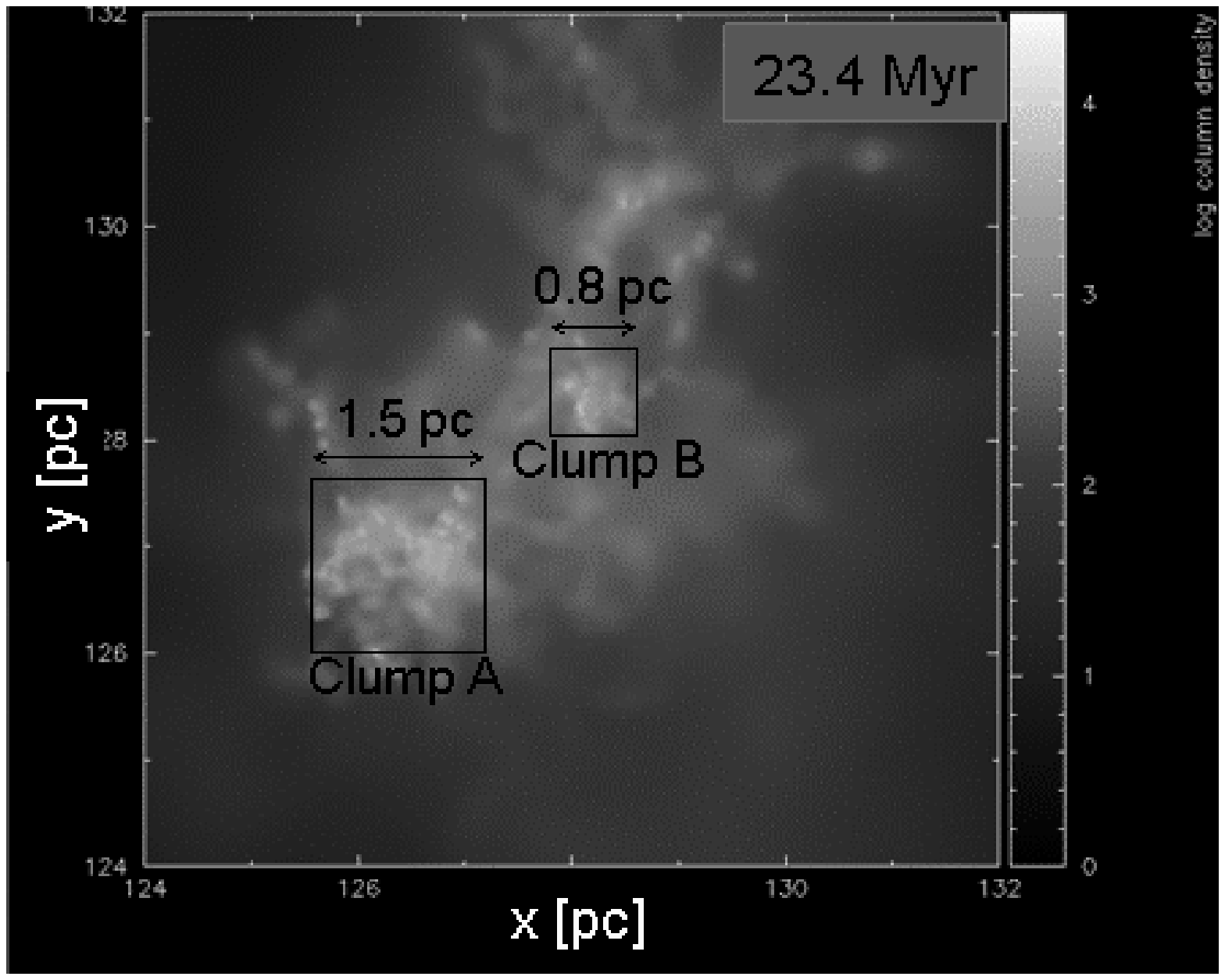}{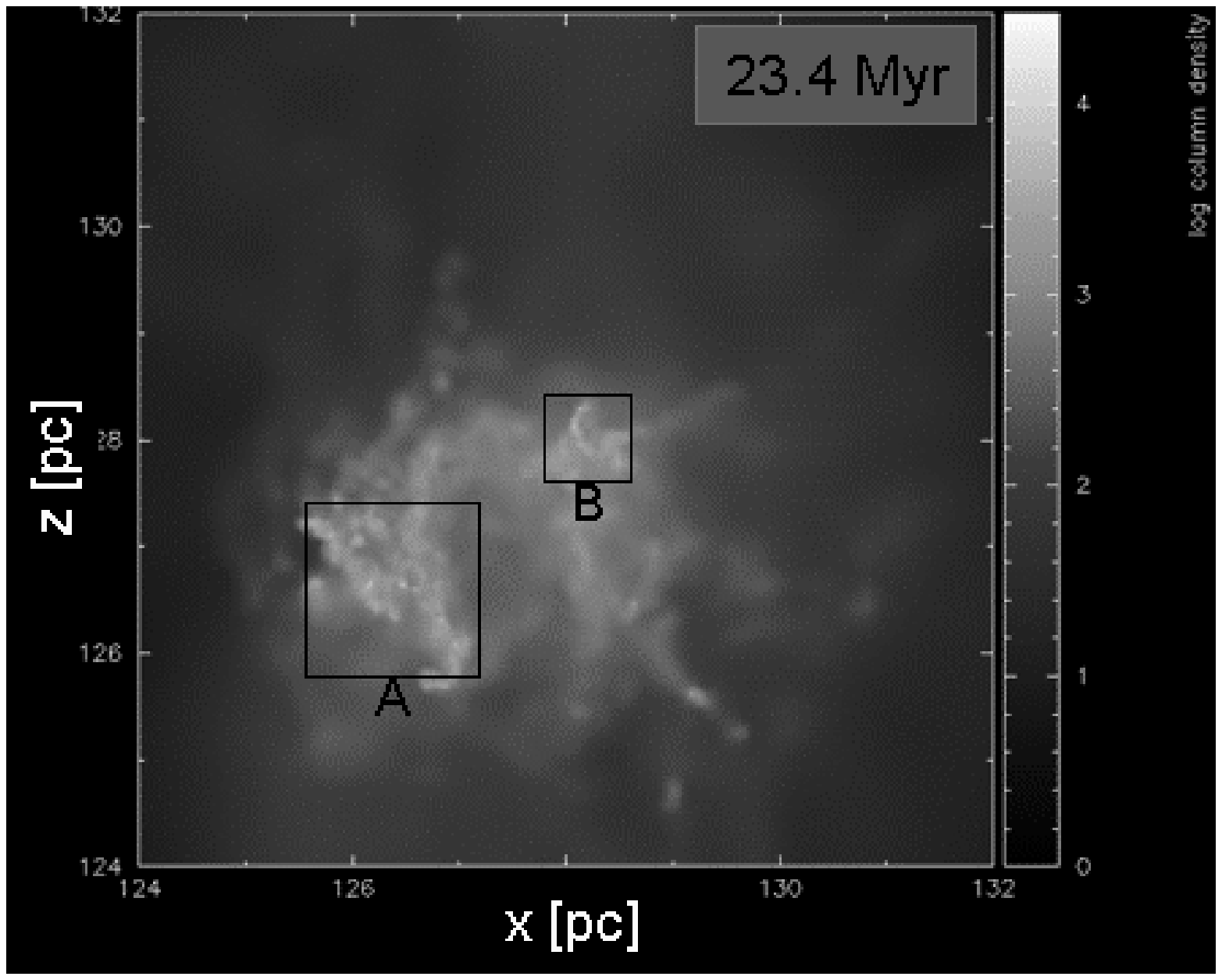}
\caption{Two views of the central 8-pc cubic region highlighted in
Fig. \ref{fig1}. {\it Left panel:} Column density integrated along the
$z$ direction. {\it Right panel:} Column density integrated along the
$y$ direction. The two squares show the regions called Clump A and Clump
B.}
\label{fig2}
\end{figure}

Figures \ref{fig2}a and \ref{fig2}b show two views of two parsec-sized clumps
within the 8-pc cloud, to we refer as ``Clump A'' and ``Clump B''. Note
that these are just cubic boxes enclosing the dense clumps, rather than
actual clumps defined by any clump-finding algorithm. Their properties
can be compared with the observed typical properties of clumps in
high-mass star forming regions. We do this by interpolating the SPH data
for the central 8-pc region into a fixed grid with a resolution of
$256^3$. Clump A, which has a size of 1.5 pc per
side, has a mass ${\cal M} = 1400 M_\odot$, a mean density of $\langle n
\rangle = 1.27 \times 10^4$ cm$^{-3}$, and a three-dimensional velocity
dispersion $\sigma = 3.6$ km s$^{-1}$. Clump B, in turn, has a linear
size of 0.8 pc, a mass ${\cal M} = 300 M_\odot$, a mean density of
$\langle n \rangle = 1.72 \times 10^4$ cm$^{-3}$, and a velocity
dispersion $\sigma = 2.8$ km s$^{-1}$. So, in general these properties
compare well with those quoted in \S \ref{sec:intro}, except perhaps for
slightly lower mean densities than typical, which can be understood
as a consequence of our usage of cubic boxes rather than clumps. The
boxes include some lower-density gas. On the other
hand, our densities fare in well with those reported for massive
starless clumps in the Cygnus X molecular complex by
\citet{Motte_etal07}.

\begin{figure}
\plotone{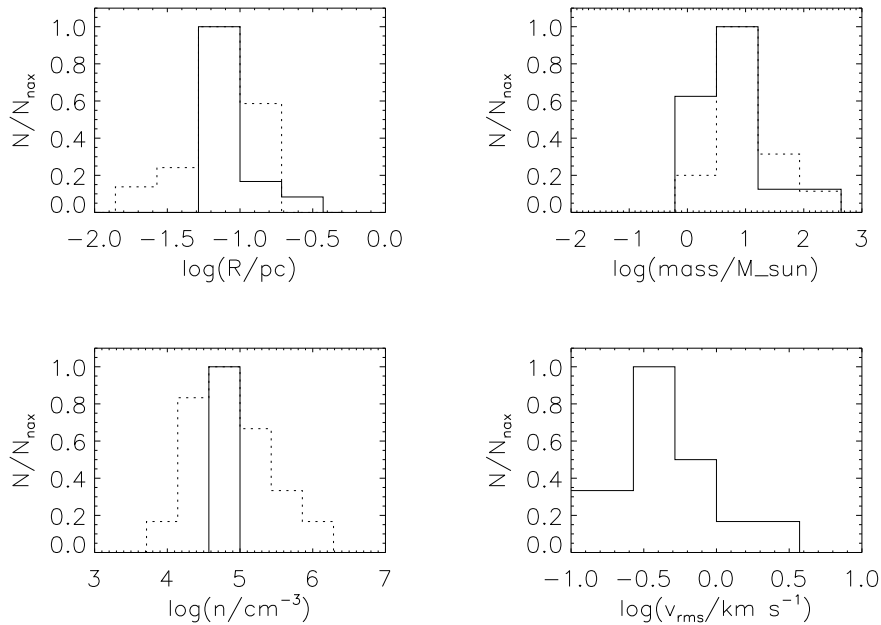}
\caption{Histograms of the size ({\it top left}), mass ({\it top
right}), mean density ({\it bottom left}) and three-dimensional velocity
dispersion ({\it bottom right}) of the cores within Clumps A and B of
the simulation ({\it solid lines}), normalized to the peak of the
histogram. {\it Dotted lines}: same for the core sample in Cygnus X
reported by \citet{Motte_etal07}, except for the velocity dispersion,
since their observations were performed in the 1.2~mm continuum. The
peaks of the distributions are seen to coincide for size, mass and mean
density. The reasons for the simulation distributions being narrower are
discussed in the text.}
\label{fig3}
\end{figure}

Within these clumps, we identify cores by applying a simple
clump-finding algorithm based on finding connected sets of grid points
whose densities are above a certain threshold. For this preliminary
analysis, we consider a single threshold $n_{\rm thr} = 5 \times 10^4$
cm$^{-3}$, leaving us with 20 cores, 14 of which have $M > 4 M_\odot$. For
each core, we measure its mass, 
mean density, and velocity dispersion, and estimate its size as $R
\approx (3 V/4 \pi)^{1/3}$, where $V$ is its volume. These properties
can be compared with those reported by recent surveys of cores in
high-mass star forming regions, such as that by \citet{Motte_etal07} for
the Cygnus X region. These authors give these data for a sample of 129
dense cores with sizes $\sim 0.1$ pc, masses 4--950 $M_\odot$, and mean
densities $\sim 10^5$ cm$^{-3}$. In order to compare their data to the
cores in our clumps, we fortuitously select the first 57 cores (those in
the first page) in their Table 1 and show the distribution of their
properties in the histograms presented in Fig.\ \ref{fig3} by the {\it
dotted} lines. Superimposed on these histograms, the {\it solid} lines
show the corresponding distributions for the 14 cores more massive than
$4 M_\odot$ in Clumps A and B of
our simulation. We see that, although the distributions of the Cygnus X
cores are in general broader than those of the cores in our simulation,
the peaks of the distributions match for size, mass, and mean density.

The fact that the distribution of densities for the cores in the
simulation is narrower than the Cygnus X one is most likely a result of
our sample being significantly smaller than that for Cygnus X, and of
our having considered only a single density threshold for defining the
cores. Our previous experience with this clump-finding method is that
the mean density of the cores found is generally within less than one
order of magnitude of the threshold density. This limitation is also the
probable cause of the absence of the small-size tail of the cores in the
simulation, since smaller cores should appear when higher density
thresholds are considered. So, in order to obtain a wider range of
densities, masses and sizes, it is necessary to consider a suite of
density thresholds. We plan to do it in the final study, to be presented
elsewhere.

%No comparison can be made for the distribution of velocity
%dispersions, since \citet{Motte_etal07}'s observations were performed in
%the radiocontinuum.

\section{Conclusions} \label{sec:conclusions}

In this contribution we have presented preliminary numerical evidence
that the physical conditions in high-mass star forming regions can arise
from global gravitational infall, with the velocity dispersions being
caused primarily by infall motions rather than random turbulence. The
evidence comes from the first study of core properties in a simulation
of the entire evolution of a molecular cloud, from its formation in the
diffuse atomic ISM to its gravitational collapse. Although the analysis
presented here is only preliminary, it is consistent with recent
suggestions, based on comparisons between simulations and observations,
that molecular clouds \citep[e.g.,][]{HB07} and clumps
\citep[e.g.][]{PHA07} may be in a state of gravitational collapse. If
confirmed, these suggestions point towards a return to the original
suggestion by \citet{GK74} that the observed linewidths in molecular
clouds are due primarily to gravitational contraction. This suggestion
was dismissed by \citet{ZP74} through the argument that this would imply
a much larger average star formation rate in the Galaxy than observed.
However, it is possible that this criticism may be overcome if magnetic
field fluctuations in the clouds imply that some parts of them are
magnetically supported so that only the non-supported parts parts of the
clouds undergo collapse \citep{HBB01, Elm07}, and the star formation
rate is then regulated by stellar feedback, with globally collapsing
motions arising only for those regions that manage to ``percolate''
through the field fluctuations and the stellar-feedback motions. We plan
to perform simulations including magnetic support and stellar feedback
in future studies to investigate this possibility.

\acknowledgements %%% Text of acknowledgements runs on after this command.
The numerical simulation was performed in the cluster at CRyA-UNAM
acquired with CONACYT grants to E.V.-S. 36571-E and 47366-F. The
visualization was produced using the SPLASH code \citep{Price07}. We
also thankfully acknowledge financial support from grants UNAM-PAPIIT 110606
to J. B.-P., and SFB 439, ``Galaxies in the Young Universe'', funded by
the German Science Foundation (DFG), to R.S.K.

%%% THE BIBLIOGRAPHY
%%%
%%% CONSULT SECTION 3 OF "INSTRUCTIONS FOR AUTHORS" FOR HOW TO USE NATBIB.
%%% AUTHORS ARE ENCOURAGED TO USE EITHER THE "THEBIBLIOGRAPY" ENVIRONMENT
%%% BY UNCOMMENTING (DELETING THE "%" SYMBOL) THE COMMANDS BELOW, OR BY
%%% USING THE BIBTEX ENVIRONMENT. TO FIND OUT WHICH IS APPLICABLE TO YOUR
%%% CONTRIBUTION, CONSULT THE VOLUME EDITORS FOR YOUR PROCEEDINGS.
%%%


\begin{thebibliography}{}

\bibitem[Audit \& Hennebelle (2005)]{AH05} Audit, E. \& Hennebelle,
P. 2005, A\&A 433, 1

\bibitem[Ballesteros-Paredes et al.(1999b)Ballesteros-Paredes, 
Hartmann, \& V\'azquez-Semadeni]{BHV99} Ballesteros-Paredes, J.,
Hartmann, L. \& V\'azquez-Semadeni, E. 1999b, ApJ 527, 285

\bibitem[Ballesteros-Paredes et al.(1999a)Ballesteros-Paredes, 
V\'azquez-Semadeni, \& Scalo]{BVS99} Ballesteros-Paredes, J.,
V\'azquez-Semadeni, E., \& Scalo, J. 1999a, ApJ, 515, 286

\bibitem[Ballesteros-Paredes et al.(2008)]{BP_etal08}
Ballesteros-Paredes, J., Gazol, A., V\'azquez-Semadeni, E., \& Kim,
J. 2008, to be submitted to MNRAS

\bibitem[Bate, Bonnell, \& Bromm(2003)]{BBB03}
Bate, M. R., Bonnell, I. A., \& Bromm, V. 2003, MNRAS 336, 705 

\bibitem[Beuther et al.(2007)]{Beuther_etal07} Beuther , H., Churchwell,
E. B., McKee, C. F., Tan, J. C. 2007, in Protostars and Planets V,
eds. B. Reipurth, D. Jewitt, and K. Keil (Tucson: University of Arizona
Press), 165

\bibitem[Elmegreen (1991)]{Elm91} Elmegreen, B. G. 1991, in The Physics
of Star Formation and Early Stellar Evolution, ed.\ C.J. Lada and
N. D. Kylafis (Dordrecht: Kluwer), 35

\bibitem[Elmegreen(2007)]{Elm07} Elmegreen, B. 2007, ApJ, 668, 1064

\bibitem[Field, Blackman, \& Keto(2008)]{FBK08}
Field, G., Blackman, E. G, Keto, E. R. 2008, MNRAS, submitted
(astro-ph/0601574v4)

\bibitem[Frisch(1995)]{Frisch95} Frisch,
U. 1995, Turbulence. The legacy of A.N. Kolmogorov (Cambridge: Cambridge
University Press) 

\bibitem[Garay \& Lizano(1999)]{GL99} Garay, G. \& Lizano, S. 1999,
PASP, 111, 1049

\bibitem[Goldreich \& Kwan(1974)]{GK74}
Goldreich, P., \& Kwan, J. 1974, ApJ 189, 441

\bibitem[Hartmann, Ballesteros-Paredes, \& Bergin (2001)]{HBB01} Hartmann, L.,
Ballesteros-Paredes, J., \& Bergin, E. A. 2001, ApJ, 562, 852

\bibitem[Hartmann \& Burkert(2007)]{HB07}
Hartmann, L. \& Burkert, A. 2007, ApJ, 654, 988

\bibitem[Heitsch et al.(2005)]{Heitsch_etal05}
Heitsch, F., Burkert, A., Hartmann, L., Slyz, A. D. \& Devriendt,
J. E. G. 2005, ApJ 633, L113

\bibitem[Heitsch et al.(2006)]{Heitsch_etal06}
Heitsch, F., Slyz, A., Devriendt, J.,  Hartmann, L., \& Burkert, A.
2006, ApJ 648, 1052

\bibitem[Heitsch, Mac Low \& Klessen(2001)]{HMK01} Heitsch, F., Mac Low,
M. M., \& Klessen, R. S. 2001, ApJ, 547, 280

\bibitem[Hennebelle \& Audit(2007)]{HA07} Hennebelle, P. \& Audit,
E. 2007, A\&A, 465, 431

\bibitem[Hunter \& Fleck(1982)]{HF82} Hunter,
J. H., Jr., \& Fleck, R. C., Jr. 1982, ApJ, 256, 505 

\bibitem[Jappsen et al.(2005)]{Jappsen_etal05} Jappsen, A.-K.,
Klessen, R.~S., Larson, R.~B., Li, Y., and Mac Low, M.-M. 2005, A\&A 435, 611

\bibitem[Kim, Ostriker \& Stone(2002)]{KOS02} Kim, W.-T., Ostriker,
E. C. \& Stone, J. M. 2002, ApJ, 581, 1080

\bibitem[Klessen, Heitsch \& Mac Low(2000)]{KHM00} Klessen, R. S.,
Heitsch, F., \& MacLow, M. M. 2000, ApJ, 535, 887

\bibitem[Koyama \& Inutsuka (2002)]{KI02} Koyama, H. \& Inutsuka,
S.-I. 2002, ApJ, 564, L97

\bibitem[Kurtz et al.(2000)]{Kurtz_etal00} Kurtz, S., Cesaroni, R.,
Churchwell, E., Hofner, P., Walmsley, C. M. 2000, in Protostars and
Planets IV, eds. V. Mannings, A. P. Boss, S. S. Russell (Tucson:
University of Arizona Press), 299

\bibitem[McKee \& Tan(2003)]{MT03}McKee, C. F., Tan, J. C. 2003, ApJ, 585, 850 

\bibitem[Motte et al.(2007)]{Motte_etal07} Motte, F., Bontemps, S.,
Schilke, P., Schneider, N., Menten, K. M., Broguière, D. 2007, A\&A,
476, 1243

\bibitem[Passot et al.(1995)Passot, V\'azquez-Semadeni \&
Pouquet]{PVP95} Passot, T., V\'azquez-Semadeni, E., \& Pouquet, A. 1995,

\bibitem[Peretto, Hennebelle \& Andr\'e(2007)]{PHA07} Peretto, N.,
Hennebelle, P., Andr\'e, P. 2007, A\&A, 464, 983 

\bibitem[Price(2007)]{Price07} Price, D. J. 2007, PASA, 24, 159

\bibitem[Springel et al.(2001)]{Springel_etal01}
Springel, V., Yoshida, N., White, S. D. M. 2001, New Astron., 6, 79

\bibitem[V\'azquez-Semadeni et al.(2007)]{VS_etal07} V\'azquez-Semadeni, E.,
G\'omez, G. C., Jappsen, A. K., Ballesteros-Paredes, J., Gonz\'alez,
R. F., \& Klessen, R. S. 2007, ApJ, 657, 870 (Paper I)

\bibitem[V\'azquez-Semadeni et al.(2008)]{VS_etal08} V\'azquez-Semadeni,
E., Gonz\'alez, R. F., Ballesteros-Paredes, J., Kim, J. 2008, MNRAS, to
be submitted

\bibitem[V\'azquez-Semadeni et al.(2005)]{VKSB05} V\'azquez-Semadeni,
E., Kim, J., Shadmehri, M. \& Ballesteros-Paredes, J. 2005, ApJ, 618, 344 

\bibitem[V\'azquez-Semadeni et al.(1995)V\'azquez-Semadeni, Passot, \&
Pouquet]{VPP95} V\'azquez-Semadeni, E., Passot, T., \& Pouquet, A. 1995,
ApJ, 441, 702

\bibitem[V\'azquez-Semadeni et al.(1996)V\'azquez-Semadeni, Passot, \&
Pouquet]{VPP96} V\'azquez-Semadeni, E., Passot, T., \& Pouquet, A. 1996,
ApJ, 473, 881 

\bibitem[V\'azquez-Semadeni et al.(2006)]{VS_etal06}
V\'azquez-Semadeni, E., Ryu, D., Passot, T., Gonz\'alez, R. F., \&
Gazol, A.., 2006, ApJ, 643, 245 

\bibitem[Vishniac(1994)]{Vishniac94} Vishniac, E. T. 1994, ApJ, 428, 186

\bibitem[Walder \& Folini(1998)]{WF98} Walder, R. \& Folini, D. 1998
A\&A, 330, L21

\bibitem[Zuckerman \& Palmer(1974)]{ZP74} 
Zuckerman, B. \& Palmer, P. 1974, ARA\&A, 12, 279

\end{thebibliography}
\end{document}